\documentclass[useAMS,usegraphicx]{mn2e}
\usepackage{rotate}
\usepackage{times}
\newif\ifAMStwofonts
\AMStwofontstrue

\def\gs{\mathrel{\hbox{\rlap{\hbox{\lower4pt\hbox{$\sim$}}}\hbox{$>$}}}}
\def\ls{\mathrel{\hbox{\rlap{\hbox{\lower4pt\hbox{$\sim$}}}\hbox{$<$}}}}

\def\Msun{M$_{\odot}$}

\def\xmm{{\it XMM-Newton}}

\def\et{{et al.\ }}
\def\mcg{{MCG--6-30-15}}
\def\iz{{I~Zw~1}}
\def\3c{{3C~273}}

\def\rg{{\thinspace r_{\rm g}}}

\def\ka{{K$\alpha$}}

\def\keV{{\rm\thinspace keV}}

\def\kpc{{\rm\thinspace kpc}}

\def\Msun{\hbox{$\rm\thinspace M_{\odot}$}}

\def\s{{\rm\thinspace s}}
\def\ks{{\rm\thinspace ks}}

\title[Distinct modes of spectral variability in \iz]
      {
A longer \xmm\ look at I Zwicky 1: 
Distinct modes of X-ray spectral variability 
      }

\author[L. C. Gallo et al.]
       {L. C. Gallo,$^{1,2}$  
        W. N. Brandt,$^3$ 
        E. Costantini$^{4,5}$ and
	A. C. Fabian$^6$ \\ 
$^{1}$ SUPA, School of Physics and Astronomy, University of St Andrews, North Haugh, St Andrews, Fife KY16 9SS \\
$^{2}$ Max-Planck-Institut f\"ur extraterrestrische Physik, Postfach 1312, 85741 Garching, Germany \\
$^{3}$ Department of Astronomy and Astrophysics, The Pennsylvania State University, 525 Davey Lab, University Park, PA 16802, USA \\
$^{4}$ SRON National Institute for Space Research Sorbonnelaan 2, 3584 CA Utrecht, The Netherlands \\
$^{5}$ Astronomical Institute, Utrecht University, P.O. Box 80000, 3508 TA
Utrecht, The Netherlands \\
$^{6}$ Institute of Astronomy, University of Cambridge, Madingley Road, Cambridge CB3 0HA\\
}
\date{Accepted. Received. }
\pagerange{\pageref{firstpage}--\pageref{lastpage}}
\pubyear{2007}
\begin{document}
\maketitle
\label{firstpage}

\begin{abstract}
The short-term spectral variability of the narrow-line Seyfert 1 galaxy 
I Zwicky 1 (\iz) as observed in an $85\ks$ \xmm\ observation is discussed in 
detail.
\iz\ shows distinct modes of variability prior to and after a flux dip in
the broad-band light curve.  Before the dip the variability can be described
as arising from changes in shape and normalisation of the spectral components.  
Only changes in normalisation are manifested after the dip.  
The change in the mode of behaviour occurs on dynamically short
timescales in \iz.
The data suggest that
the accretion-disc corona in \iz\ could have two components that are
co-existing.  The first, a uniform, physically diffuse plasma responsible for
the ``typical'' long-term (e.g. years) behaviour; and a second compact,
centrally located component causing the rapid flux and spectral changes.
This compact component could be the base of a short or aborted jet as
sometimes proposed for radio-quiet active galaxies.
Modelling of the 
average and time-resolved rms
spectra demonstrate that a blurred Compton-reflection model can describe the
spectral variability if we allow for pivoting of the continuum component prior
to the dip. 
\end{abstract}

\begin{keywords}
galaxies: active -- 
galaxies: nuclei -- 
quasars: individual: \iz\  -- 
X-ray: galaxies 
\end{keywords}


\section{Introduction}
\label{sect:intro}
There is strong evidence that the accretion processes in Galactic black
holes (GBHs) and active galactic nuclei (AGNs) are similar within
some scaling relation (e.g. Merloni \et 2003; McHardy \et 2006).
It would then follow that the various states and types of spectral
variability seen in GBHs (see
Remillard \& McClintock 2006 for a recent review) should also be present
in AGNs.  
Here we study a spectral variability mode change in the AGN
I~Zwicky~1 (\iz; $z=0.0611$).

In the standard model
a transition from one state to another in GBHs is normally attributed to 
a change in the geometry of the accretion flow and thus should occur on viscous
timescales (e.g. Esin \et 1998).  In the high-soft state, a
standard accretion disc (Shakura \& Sunyaev 1973) extends down to the last 
stable orbit and dominates the X-ray emission.  In the low-hard 
state the standard disc is truncated at a few 100 gravitational radii
($\rg=GM/c^2$) and hard, non-thermal emission from a low-efficiency
accretion flow dominates.
Identifying similar behaviour in $\sim10^7\Msun$ AGNs is difficult as the 
characteristic timescales are $10^6$ times longer than for a $10\Msun$ 
GBH.

The low-hard state is also associated with radio emission from a jet, which 
dissipates as the GBH enters the high-soft state (Gallo \et 2003).  This has
led to the concept that much of the X-ray emission is from the base of
a jet (e.g. Markoff, Falcke \& Fender 2001; Ghisellini, Haardt \& Matt 2004).  
Moreover,
Belloni \et (1997) demonstrate that ``blobs'' of radio emission observed from  
the Galactic black
hole GRS~1915+105 could be the inner accretion disc that is being evacuated
from the system, giving rise to a truncated disc.  
On the other hand, Vadawale \et (2003) show that this situation can also be
realised if instead it is coronal material (i.e. hot plasma) that is ejected
while the inner disc remains intact.
In fact, Miller \et (2006) show evidence of a broad iron line in the spectrum
of GX~339--4 during a low-hard state indicating that an optically thick 
disc is present near the innermost stable circular orbit, in apparent
disagreement with the standard model for state transitions.
Therefore, some state changes may be associated with changes in the inner 
magnetic 
structure of the disc, which could occur faster than viscous timescales,
rather than from modification of the accretion flow. 

GBHs also exhibit much more rapid types of spectral and flux variations
(e.g. McClintock \& Remillard 2003).  For
example, in Greiner \et (1996) GRS~1915+105 is shown to undergo 
fast oscillations of various amplitudes on timescales down to a few seconds.
Catching a supermassive black hole exhibiting an analogous event over the
course of an X-ray mission is likely, but even these events would occur on
timescales longer than a typical \xmm\ AGN observation.

The narrow-line Seyfert 1 galaxy (NLS1) \iz\
was observed with \xmm\
(Jansen \et 2001) at two epochs separated by approximately three years (Gallo
\et 2004a, 2007; hereafter G04 and G07, respectively).  In G07 the mean
spectral and timing properties of the 2005 observation were presented and
changes in the spectral shape and intrinsic absorption (neutral and ionised)
since the 2002 observation were reported (see Costantini \et (2007) for
a detailed presentation of the ionised absorber).  Here we focus on the
rapid spectral variability during the 2005 observation.  Details of the
observation and data processing are found in G07.  

The black-hole mass ($M$) in \iz\ is
log($M$/\Msun)$=7.441^{+0.093}_{-0.119}$ (Vestergaard \& Peterson 2006) and
has a corresponding orbital period of $<3000s$ at $\sim2\rg$ ($\rg=GM/c^2$). 
The short-term variability
exhibited by \iz\ could be portraying an observable state transition in an AGN.
The change in the mode of behaviour in \iz\ is occurring on dynamically short
timescales.  Such variations would correspond to a few ms in GBHs
and likely be unobservable as they 
would be too short.  When normalised per unit crossing time, the count rates in
AGNs are orders of magnitude larger than in GBHs.


\section{The sharp flux dip in the 2005 X-ray light curve}
\label{sect:xlc}

\begin{figure}
\rotatebox{270}
{\scalebox{0.32}{\includegraphics{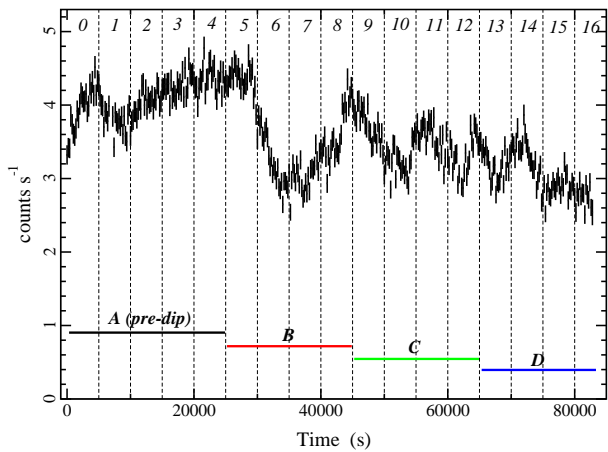}}}
\caption{The 2005 $0.2-12\keV$ EPIC pn light curve in $200\s$ bins.  
The start time corresponds to the beginning of the pn exposure.
The vertical dashed lines identify $5\ks$ blocks of the observation and are
numbered from $0-16$.  Note that block 16 is not a full $5\ks$ in length.
The entire light curve is also separated into 4 segments, each consisting
of $\sim20-25\ks$ of data, and labeled Seg A--D.  Seg A comprises data
prior to the sharp dip that starts at about $25\ks$. 
}
\label{fig:xlc}
\end{figure}

During the 2002 observation of \iz\ the most notable feature in the 
light curve was a 
modest X-ray flare that was concentrated at high energies ($E\gs3\keV$) (G04).
In the 2005 light curve, \iz\ displays another distinct feature; this time a 
sharp flux dip occurs about $25\ks$ into the observation (Fig.~\ref{fig:xlc}).

Though it is the most distinct characteristic in the light curve, the dip in 
count rates is a relatively modest event compared to the X-ray variability
seen in some NLS1s (e.g. Boller \et 1997; Brandt \et 1999).
However, of more interest is the clear change in behaviour that is
associated with the dip.  G07 already presented
the onset of a time-lag between X-ray energy bands that coincided with the
dip.  Here, the focus will be on the different spectral variability that is
exhibited during the pre- and post-dip
phases.  As will be demonstrated the flux dip represents a
transition in the spectral behaviour of the AGN.

In order to examine the time-resolved behaviour of \iz, the data were divided as
shown in Fig.~\ref{fig:xlc}.  Firstly, the data were separated into 17
blocks marked $0-16$ corresponding to $5\ks$ intervals.  As the live-time of the
pn CCD in small-window mode is $71$ per cent, the net exposure in each 
$5\ks$-interval is $\sim3.6\ks$.  In block 16 the CCD was not active for the
entire $5\ks$ resulting in limited data and poorer signal than in the other 
blocks.
Consequently, this block is omitted in the analysis if the shortage of data 
limits the constraints on model parameters.
Secondly, the data
were separated into 4 segments (labeled A--D), composed of 4 or 5 sequential
blocks each,
and identifying regions of particular interest.  For example, Segment A constitutes
blocks $0-4$, which includes all the data prior to the flux dip at $25\ks$.  
Segment B includes blocks $5-8$, which consist of data during the dip. 
Segments C and D constitute post-dip data.

\section{Pre- and post-dip: Distinct modes of variability}
\label{sect:spv}

\subsection{Hardness-ratio analysis}

\begin{figure*}
\begin{minipage}[]{0.45\hsize}
\scalebox{0.32}{\includegraphics[angle=270]{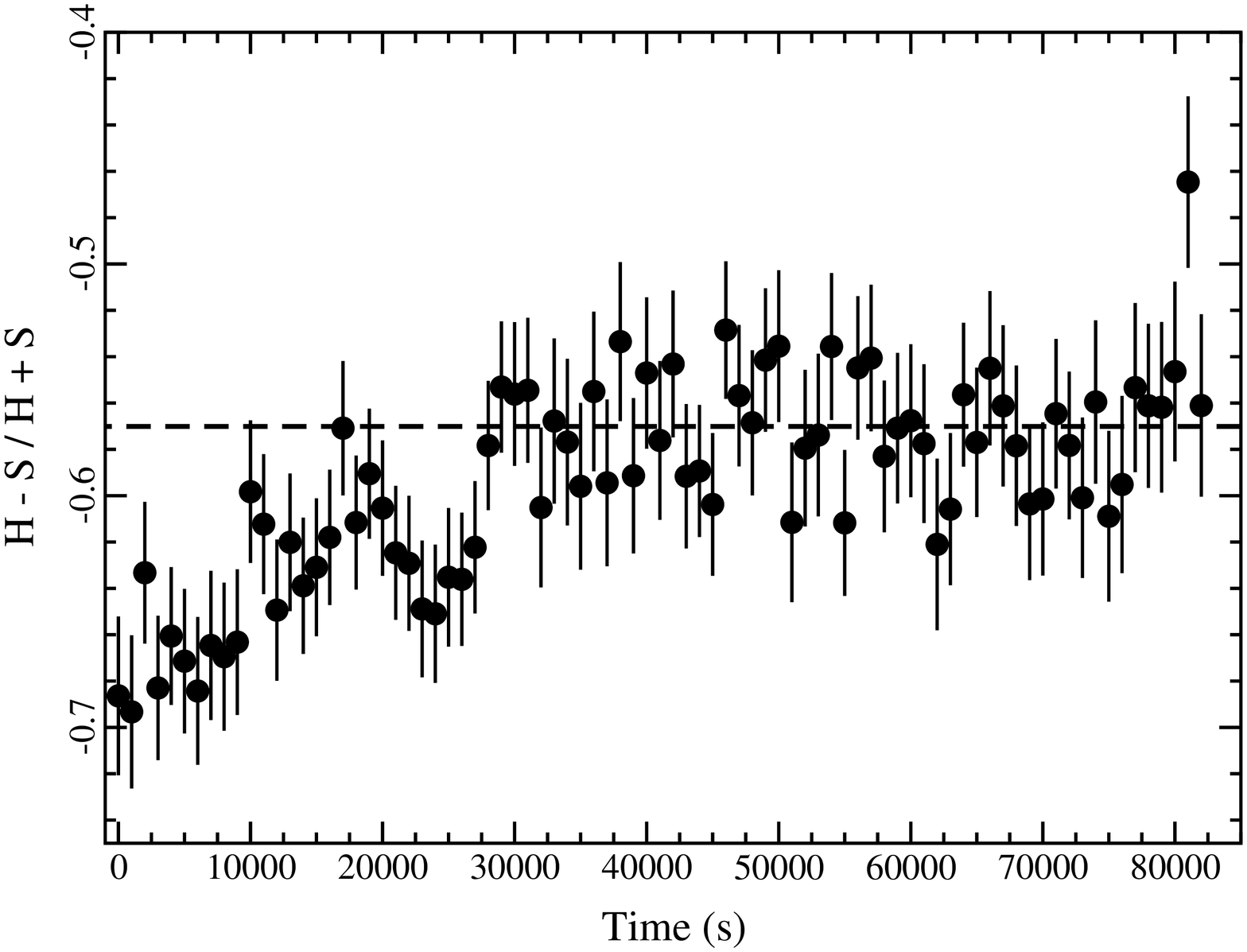}}
\end{minipage}
\hfill
\begin{minipage}[]{0.45\hsize}
\scalebox{0.32}{\includegraphics[angle=270]{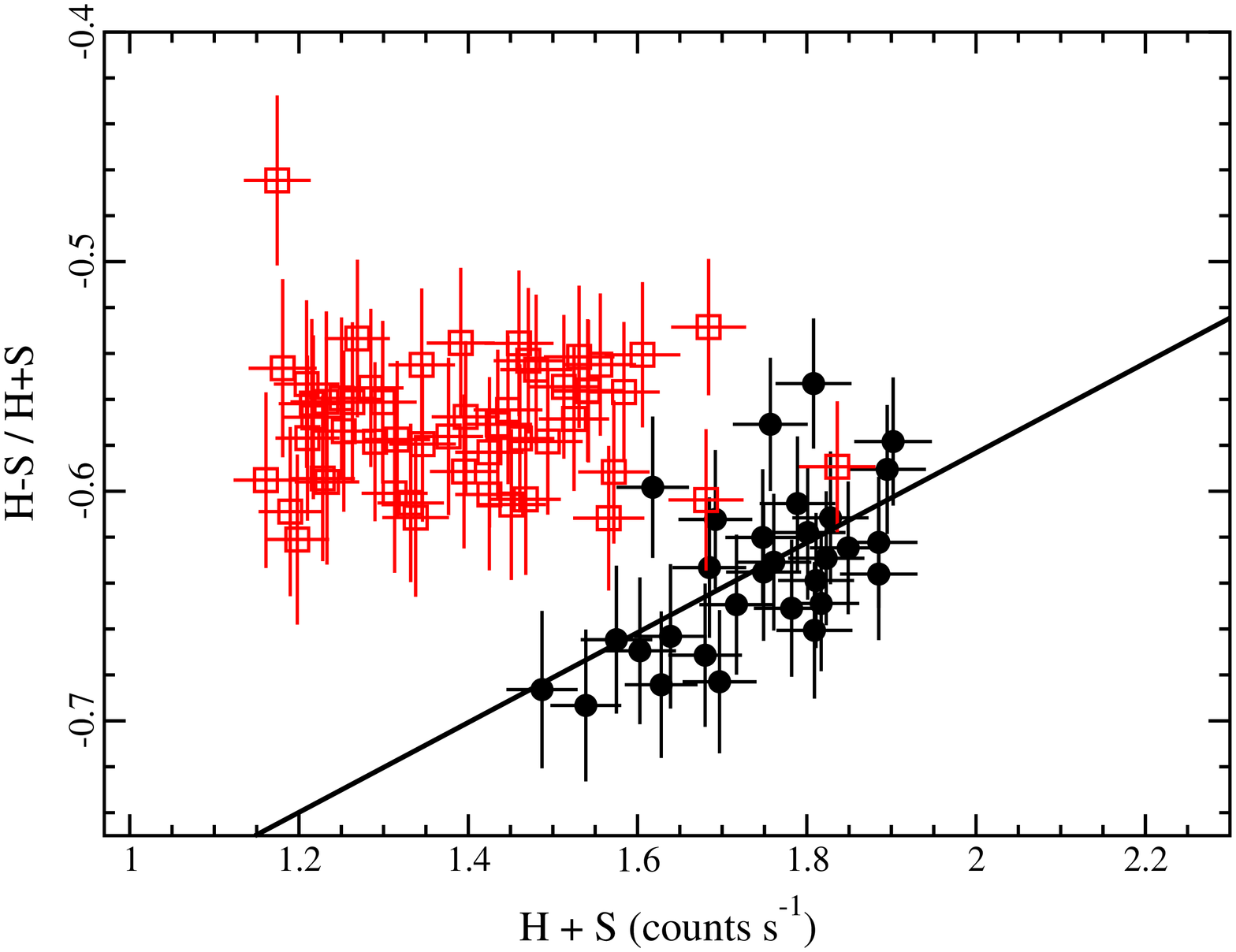}}
\end{minipage}
\caption
{\label{fig:hr}
Left panel:  A typical hardness-ratio variability curve when comparing
low- and high-energy bands.  In this case $H=2.5-12\keV$ and $S=0.5-1\keV$.  
The dashed
line marks the average hardness ratio after $30\ks$, the time corresponding
to the approximate onset of the flux dip in the light curve.  There is spectral
variability (spectral hardening) prior to this time, but negligible spectral
variability afterwards.
Right panel: Hardness ratio as a function of count rate.  The data prior
to the dip in the light curve (black dots) show a correlation between
hardness ratio and count rate (shown by the best-fit solid line).
The data after the flux dip (red squares) show no flux dependency.
}
\end{figure*}

There are several lines of evidence demonstrating distinct spectral behaviour 
prior to and after the flux dip at $\sim25\ks$.
Hardness-ratio ($HR=(H-S)/(H+S)$, where $H$ and $S$ are the count rates in the 
hard and soft bands, respectively) variability curves, such as that shown in 
Fig.~\ref{fig:hr} (left panel, $H=2.5-12\keV$ and $S=0.5-1\keV$), 
indicate that prior to the flux dip spectral 
variability (primarily spectral hardening) was taking place.
After the dip, the $HR$ curve is consistent with a
constant indicating no significant time-dependent spectral variability.
Likewise prior to the dip (Segment A), the spectral variability showed flux 
dependency
(spectral hardening with increasing count rate; right panel of Fig.~\ref{fig:hr}), 
which was not evident after the dip.

\subsection{Flux-flux analysis}
\begin{figure*}
\begin{minipage}[]{0.45\hsize}
\scalebox{0.32}{\includegraphics[angle=270]{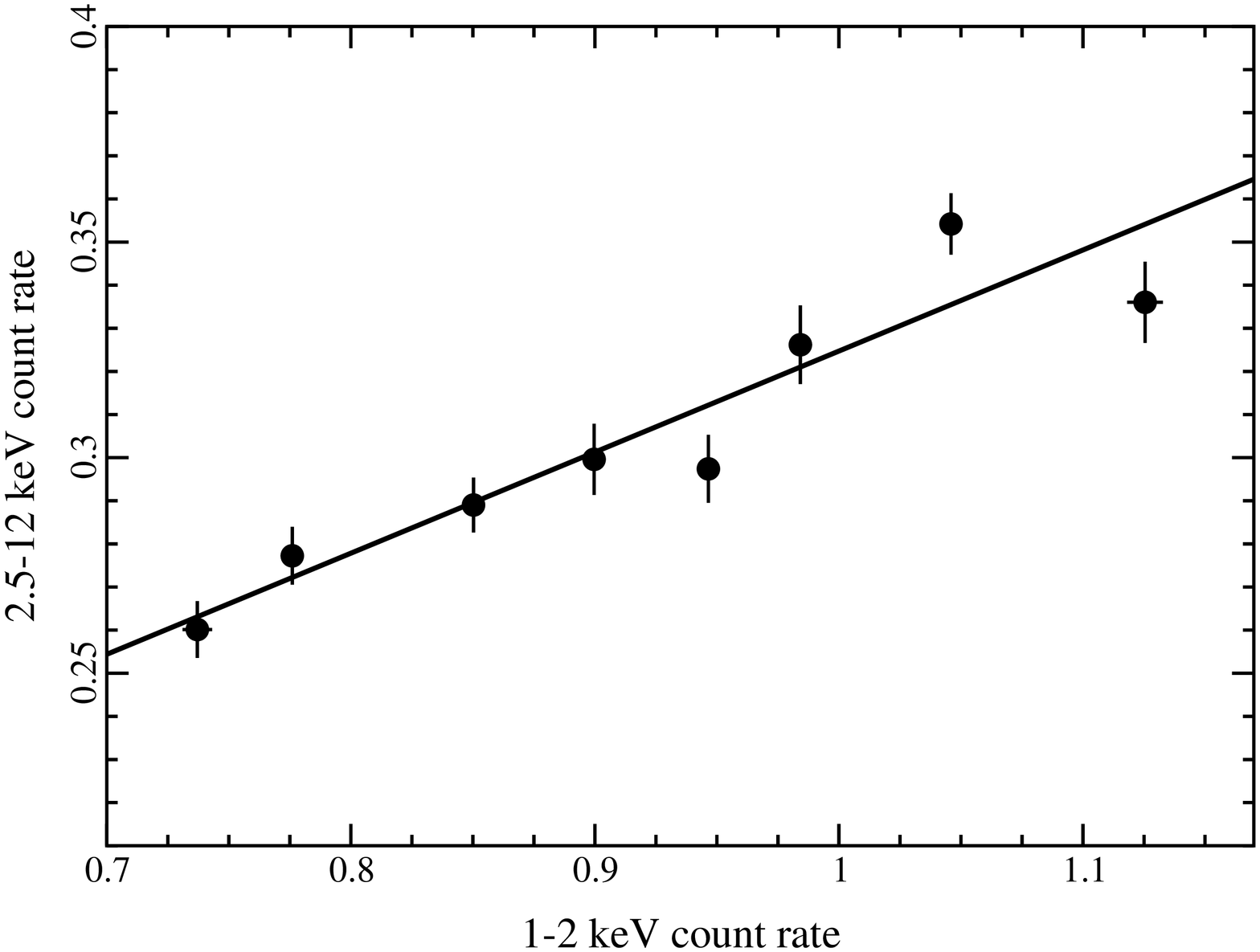}}
\end{minipage}
\hfill
\begin{minipage}[]{0.45\hsize}
\scalebox{0.32}{\includegraphics[angle=270]{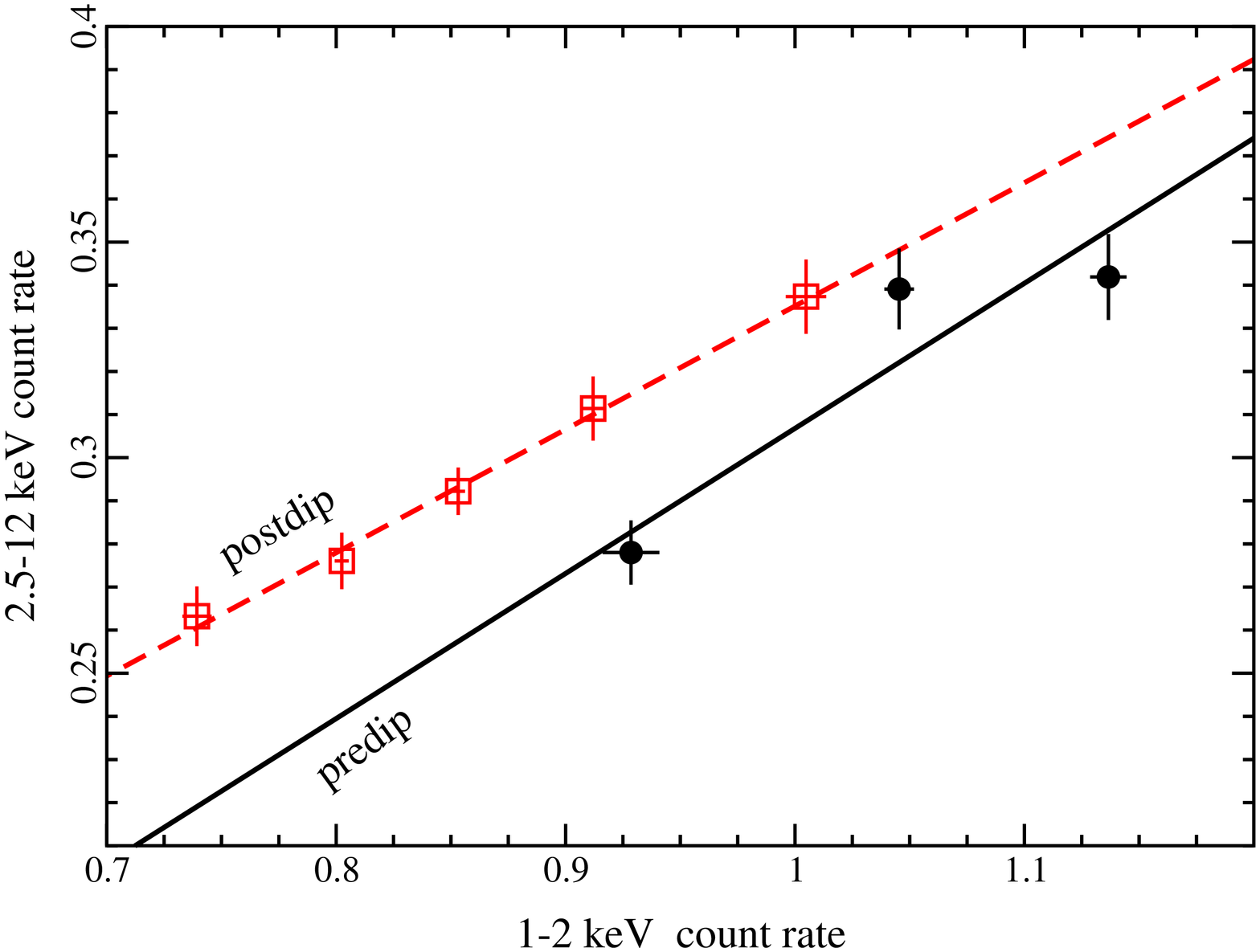}}
\end{minipage}
\caption
{\label{fig:ff05}   
Binned flux-flux plots during 2005.  Left panel: The data from the entire
observation are
not represented well with a linear fit ($\chi^2_{\nu}=1.91$).
Right panel: Creating a flux-flux diagram for the pre- and post-dip data
indicates that each data set can be described with a linear fit, but with
a different y-intercept before and after the dip.
}
\end{figure*}
Analysing the correlation between variations in two energy bands can reveal
different modes of variability (e.g. Taylor \et 2003).
For example, if the binned flux-flux plot (ff-plot)
(see Taylor \et for a complete description) can be described with a linear model 
then the variability can be associated with variations in the normalisation
of a spectral component with a constant shape.

An ff-plot constructed from the 2005 observation is shown in Fig.~\ref{fig:ff05}
(left panel).
The plot itself was binned as described in Taylor \et (2003) to overcome
intrinsic scatter in the correlation between the two energy bands.
A linear fit to the data is statistically unacceptable ($\chi^2_{\nu}=1.91$).
However, when the ff-plot for the pre- and post-dip data are considered 
separately the fitting results differ (right panel Fig.~\ref{fig:ff05}).
In this case, a line adequately fits the data from both time domains.
The data in the pre-dip phase are obviously limited so the linear fit is not
unique; however what is important is that the relation between the
two bands is different (i.e. different y-intercept) during the two
phases indicating
that the shape and mode of variability is different before and after
the flux dip.

\begin{figure}
\rotatebox{270}
{\scalebox{0.32}{\includegraphics{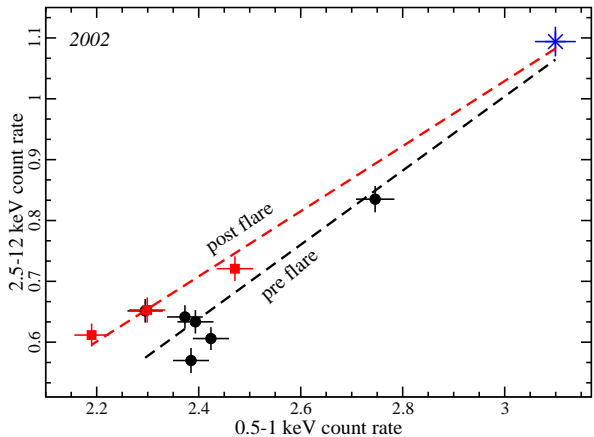}}}
\caption{A flux-flux relation during the short 2002 observation.  The plot
is not binned due to limited data.  The dashed-lines identify the best-fit
linear relation to the pre- and post-flare data (as identified in the plot).
The apparent evolution is similar to that seen in 2005 (Fig.~\ref{fig:ff05}).
}
\label{fig:ff02}
\end{figure}

Has this behaviour been seen in \iz\ before?
Fig. 9 in G02 suggests this may be true as the spectrum of \iz\ appears
softer after the hard X-ray flare.
An ff-plot was created from the 2002 data and is presented in
Fig.~\ref{fig:ff02}.
As the observation was short ($\sim20\ks$), the ff-plot is not binned as was
done in the 2005 analysis (Fig.~\ref{fig:ff05}).
What has been done here, is that a linear relation has been calculated for the 
data prior to and after the aforementioned hard X-ray flare.  
The linear relation prior to the flare ($y=-0.824+0.609x$) is slightly different 
than the relation after the flare ($y=-0.576+0.535x$). Though not highly significant
on statistical grounds, the trend in the behaviour seen in 2002 is consistent with
that seen in 2005.  That is, ``dramatic'' events in the light curve are 
consequences of, or actuate, changes in spectral behaviour.

\subsection{Flux-dependent variability}

\begin{figure}
\rotatebox{270}
{\scalebox{0.32}{\includegraphics{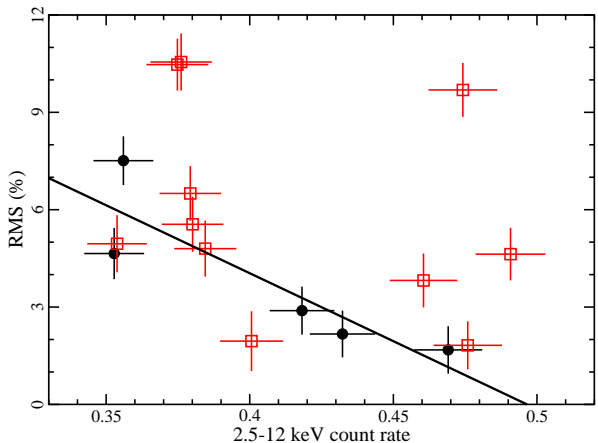}}}
\caption{The average fractional variability amplitude in each $5\ks$ block is
plotted against the corresponding $2.5-12\keV$ count rate.  An anti-correlation
in the pre-dip data (black dots) is apparent while the post-dip data (open
squares) are scattered.
}
\label{fig:crfvar}
\end{figure}

The fractional variability amplitude (e.g. Edelson \et 2002) was measured
in each $5\ks$ block and plotted against the corresponding $2.5-12\keV$
count rate (Fig.~\ref{fig:crfvar}).
Over the course of the entire observation there is no relation between
the two parameters, but if one considers the pre- and post-dip data
separately there does appear to be an anti-correlation between count rate
and fractional variability prior to the dip.  Fig.~\ref{fig:crfvar} is
shown for illustration of the potential difference prior to and after the dip.
On its own the possible difference in the behaviour is not significant,
but it appears interesting in conjunction with other lines of evidence
presented in this section.

\subsection{Time-resolved spectral fits}
\label{sect:trsf}
To this point the dual-mode spectral variability has been presented in
a completely model-independent manner, utilising only count rates and ratios
to examine spectral differences.  Here we examine the variability further by
applying simple models.

\begin{figure}
\rotatebox{270}
{\scalebox{0.32}{\includegraphics{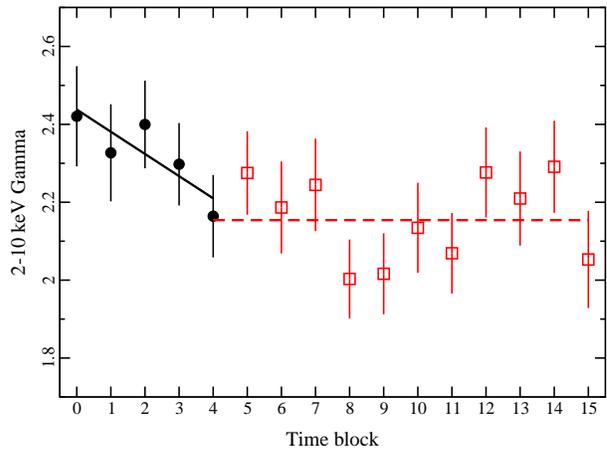}}}
\caption{The photon index as measured when fitting the $2-10\keV$
spectrum in each block with a power law (excluding the $5.9-7.1\keV$ band)
The abscissa shows the block number.  The corresponding start time of each
block can be obtained by multiplying the block number by $5000\s$.
The best-fit lines to the pre-dip (solid line) and post-dip (dashed line) 
indicate spectral hardening in the pre-dip stage and a constant photon index in 
the post-dip.
}
\label{fig:pov}
\end{figure}

The $2-10\keV$ band is dominated by a power law with an average photon index of
$\Gamma\approx2.15$ (G07).  Between $6-7\keV$ a broad feature was detected,
which is apparently variable in flux but not in energy or width (G07).
Each $5\ks$ block was modelled with a single power law in the $2-10\keV$
band while excluding the $5.9-7.1\keV$ range.  The fits were statistically
acceptable in all cases (i.e. $\chi^2_{\nu}\approx1$), and the photon index 
in each block was recorded and plotted 
(Fig.~\ref{fig:pov}).
\begin{figure}
\rotatebox{270}
{\scalebox{0.32}{\includegraphics{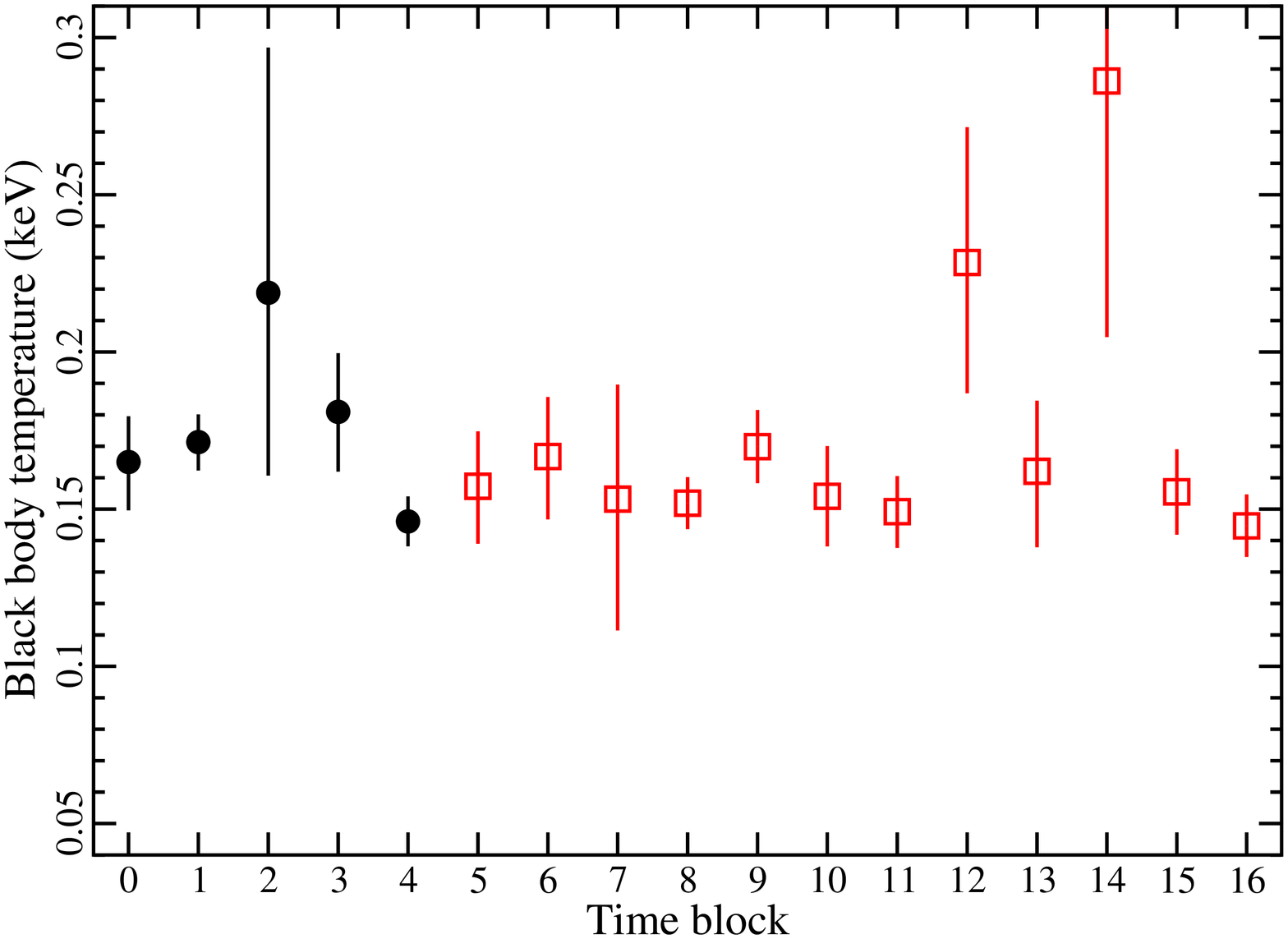}}
\scalebox{0.32}{\includegraphics{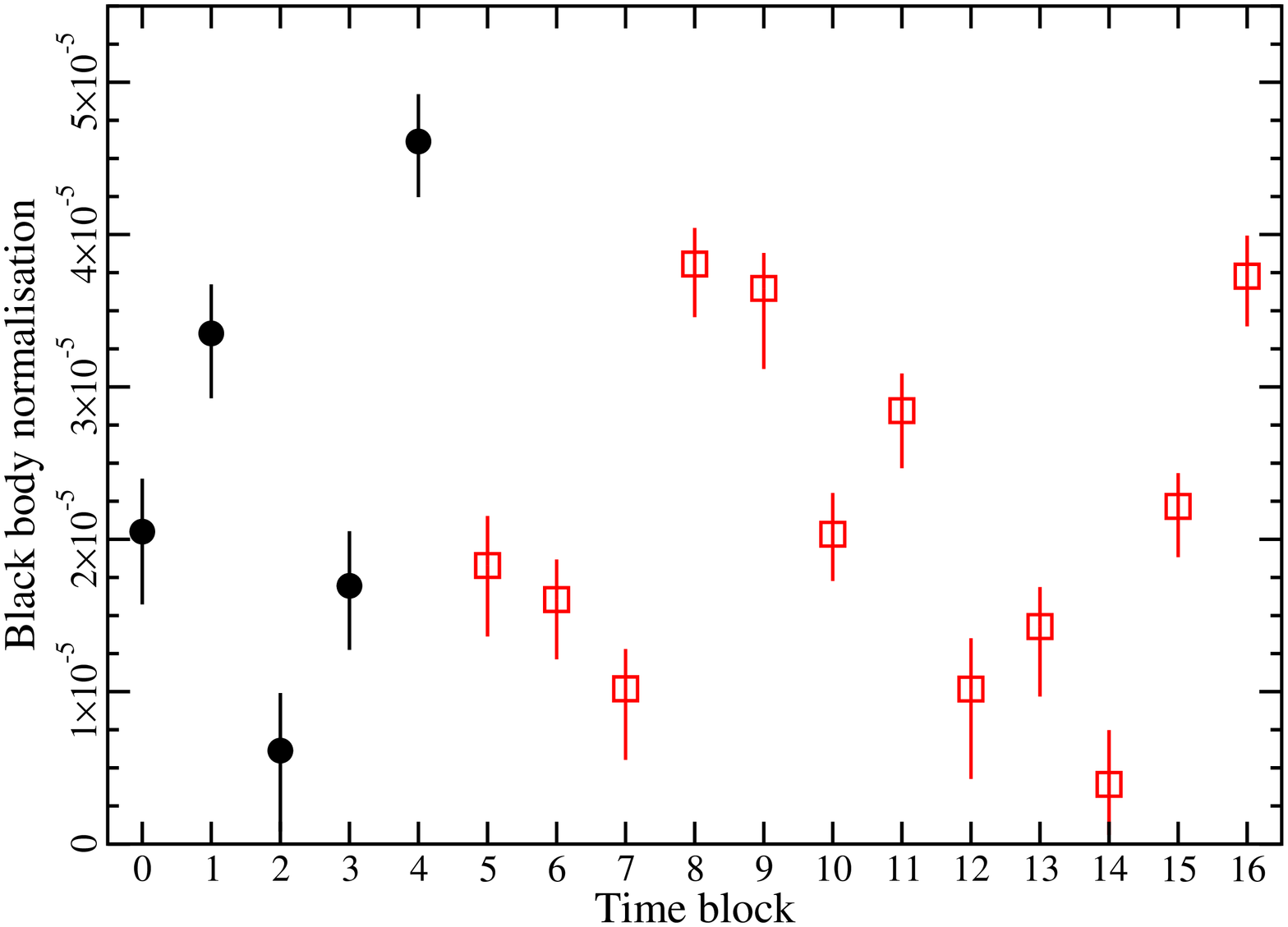}}}
\caption{A blackbody plus power law model (modified by constant neutral and
ionised absorption, see G07) is applied to each $5\ks$ block to examine 
variability in the second continuum component.
The shape of the second component (i.e. the blackbody temperature in this case)
appears constant throughout the observation (top panel).  On the other hand,
the normalisation of the component seems variable (lower panel).
The abscissa is the block number and is described in the text and in
Fig.~\ref{fig:pov}.
}
\label{fig:bbt}
\end{figure}
The average photon index is steeper during Segment A (pre-dip stage)
and appears to flatten with time.
The best fits to the pre-dip and post-data data clearly support this claim as
the post-dip data are consistent with a constant, while the pre-dip fit has
a negative slope (spectral hardening).
Even though within uncertainties the changes in photon index during the pre-dip 
phase are modest, the trend in the figure suggests pivot-like behaviour of the
power law.  The changes in the spectral shape could explain the difficulties 
encountered by G07 in fitting the average high-energy spectrum with a 
single power law.

G07 demonstrated that a second low-energy component was necessary to fit the 
broad-band
continuum of \iz, though its origin was unclear.  To examine if this second
component varied the $2-10\keV$ power law fit in each block was
extrapolated to $0.3\keV$ and a blackbody  component was added to the model.  

In addition,
the broad-band spectra were modified by neutral and ionised absorption, which
were constant over the observation,
as described in G07 and Costantini \et (2007).
The advantage of using a blackbody rather than a second power law is that the
low-energy variability could be examined empirically without altering the
fit at high energies.  However, in Section 4 we will test the variability
using spectral models with a more physical interpretation.
In all cases, the additional blackbody component improved the broad-band fit
over a single power law.  The shape (i.e. temperature) of the blackbody
in the pn bandpass did not appear notably variable,
but the normalisation did appear to change in an erratic manner
(Fig.~\ref{fig:bbt}).

\section{Modelling energy-dependent spectral variability}
\label{sect:rms}


A spectrum portraying the degree of variability in different energy bins 
(i.e. an rms spectrum; e.g. Edelson \et 2002) can offer insight into the
origin of the variations and the nature of the continuum in AGNs and GBHs 
(e.g. Gilfanov \et 2003; Fabian \et 2004;
Zdziarski 2005; Gierlinski \& Done 2006).  On relatively short 
timescales (those within a typical \xmm\ observation) the rms spectra of 
AGN usually have a concave-down shape peaking between $0.8-2\keV$ (e.g. Fabian
\et 2002, Markowitz \et 2006).
The most extreme shapes, in terms of amplitude range, are usually seen in NLS1
(e.g. Gallo \et 2004b,c).  However, the rms spectrum of \iz\ during the 2002 
observation
was unlike these, and it even displayed different shapes depending on whether
the aforementioned hard X-ray flare was included in the calculation or not 
(see figure 6 of G02).

\begin{figure*}
\begin{minipage}[]{0.45\hsize}
\scalebox{0.32}{\includegraphics[angle=270]{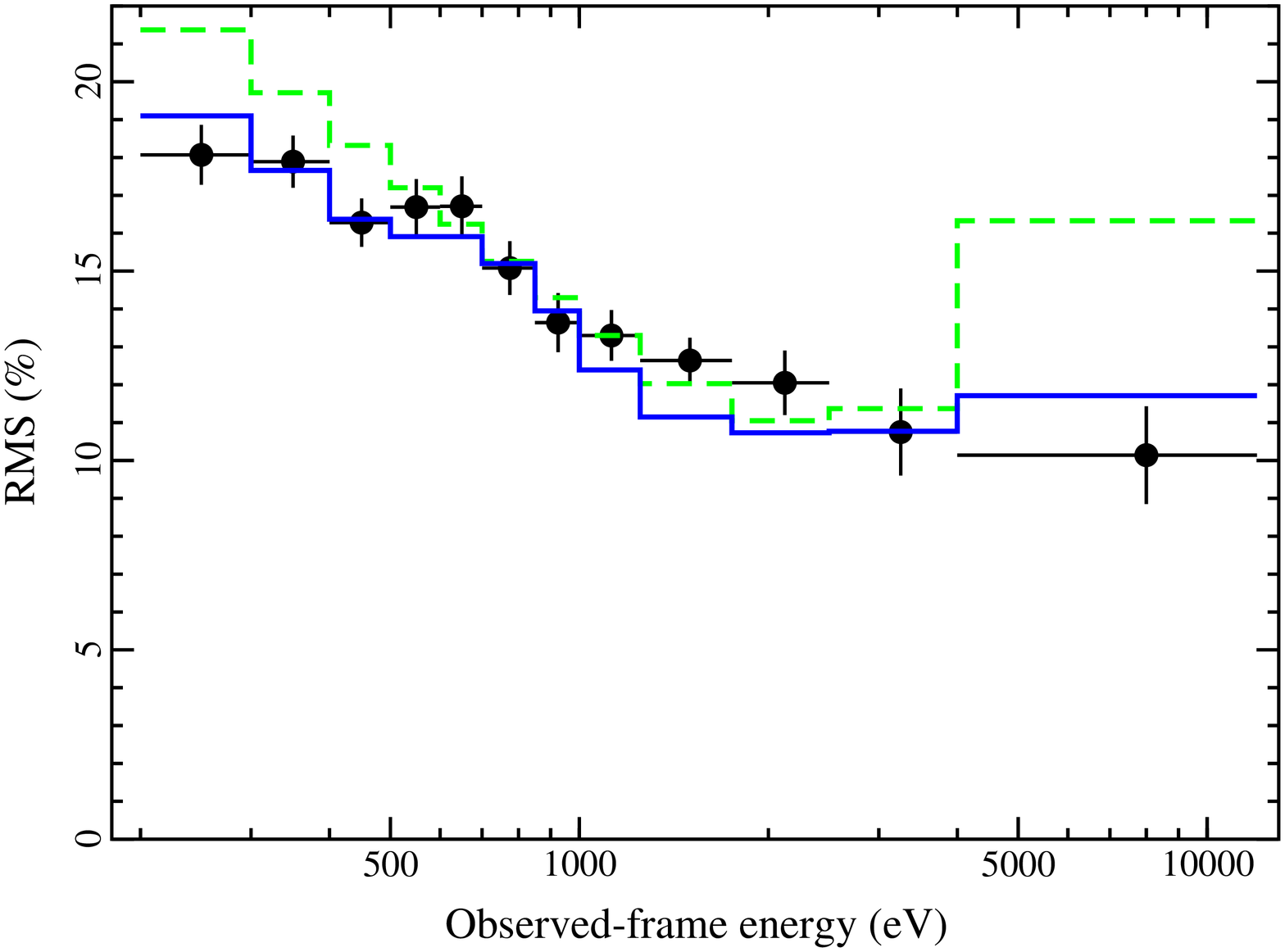}}
\end{minipage}
\hfill
\begin{minipage}[]{0.45\hsize}
\scalebox{0.32}{\includegraphics[angle=270]{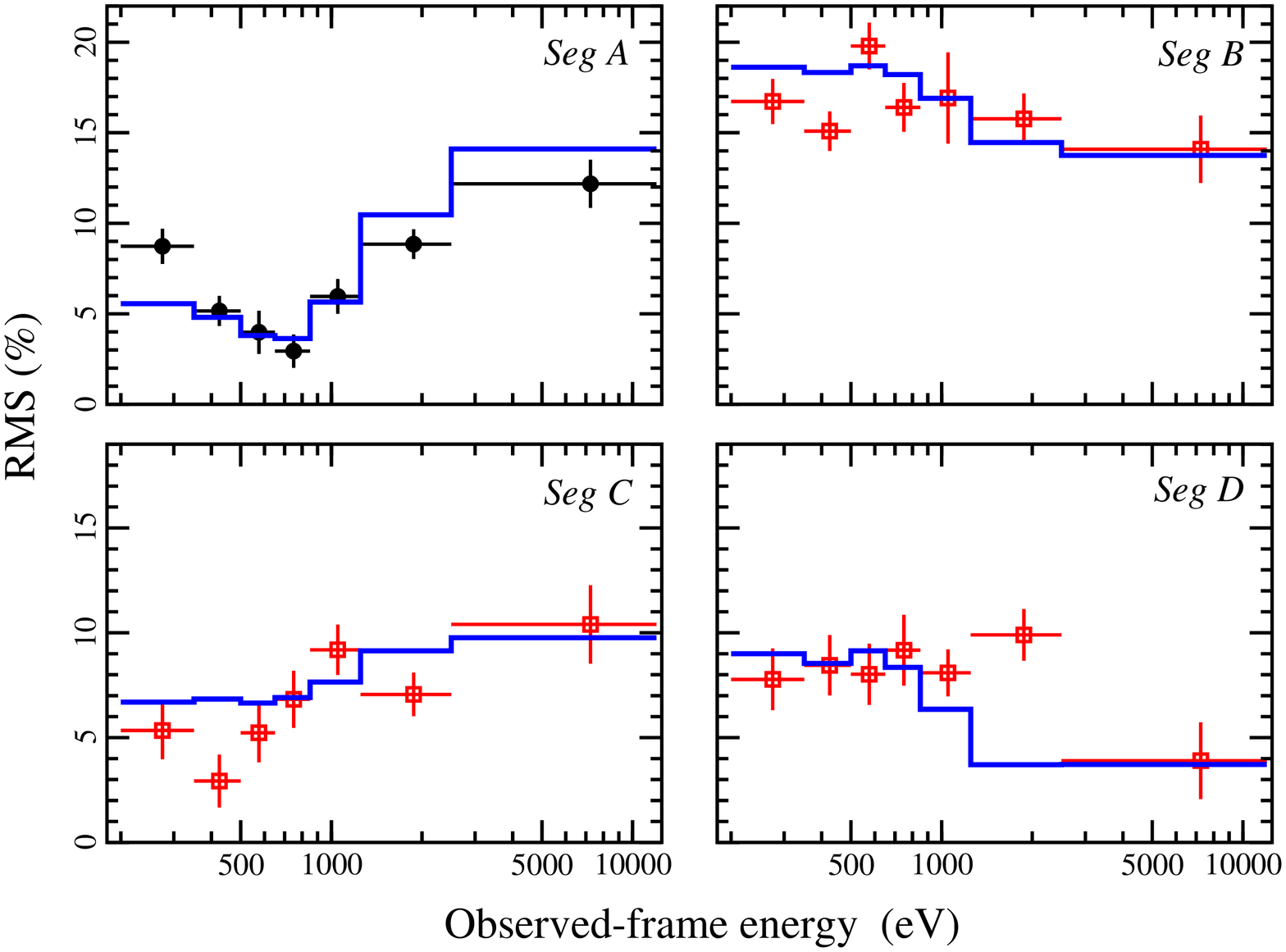}}
\end{minipage}
\caption
{\label{fig:rms}
Left panel: The average rms spectrum over the entire observation (black circles).
The dashed curve is a simulation of the rms spectrum assuming the spectrum
of \iz\ is described by a double power law.  The solid line is a simulation
of the rms spectrum assuming the spectrum is described by a blurred ionised
reflector (see Sect.~\ref{sect:rms} for details).
Right panel: An rms spectrum is calculated for each $\sim20\ks$ segment as 
indicated in the top right of each panel.  The spectrum appears distinctly
different in Segment A compared to the other three segments.  The
solid curve is a simulation of the spectra assuming the blurred ionised
reflector model in each segment (see Sect.~\ref{sect:rms} for details).
}
\end{figure*}

To ensure that the Poissonian noise was negligible compared to the rms value,
the 2005 rms spectrum was constructed with at least 
$400$ counts in each energy band and with light
curves in $1000\s$ bins.  Consequently, the uncertainties were calculated
following Ponti \et (2004).
The average rms spectrum of \iz\ at the 2005 epoch (Fig.~\ref{fig:rms} left 
panel) is unlike the one seen in
2002.  This time, the amplitude of the variations tends to increase toward
lower energies.

As the rms spectrum in 2002 appeared to vary within the observation (i.e.
it exhibited non-stationary behaviour), we searched for similar behaviour
in 2005.  An rms spectrum was created for each segment (A--D) labeled in 
Fig.~\ref{fig:xlc}.  The duration of each segment was approximately $20\ks$
and thus they could be compared to each other.  Similar to the 2002 observation
the rms spectra in the different segments appear different in shape
(Fig.~\ref{fig:rms} right panel).  In particular, the spectrum in
Segment A is significantly different than in the other time-segments, which  
are not completely dissimilar.

In establishing a potential model for the rms spectrum in Fig.~\ref{fig:rms}
we are provided with insight from Section~\ref{sect:trsf}. 
What we learned there was that of the two continuum components that
make up the average spectrum both vary in normalisation.  However,
only the power law component varies in shape and only during the pre-dip phase.

We considered first a double power law model, found to fit
the average spectrum in G07 reasonably well.  This could be interpreted
as due to multiple coronae or a continuous, multi-zone corona where 
both power laws are produced by
inverse Compton scattering of low-energy photons.  If the time lag between 
energy bands reported in G07 indicates physical separation of these coronae,
then the harder power law is produced in a corona that is more distant from the
seed photons (e.g. the accretion disc) than the corona producing the softer
power law.  Consequently, we assumed that the shape of the harder power law
remained constant over the observation, but the normalisation was allowed to vary.
The softer power law component was permitted to vary in normalisation throughout
the observation, but its photon index was only free to vary during Segment A.
The spectrum in each block was fitted with this model accordingly, and 
the predicted rms spectrum was constructed (plotted over the observed rms
spectrum in Fig.~\ref{fig:rms}).  
The rms spectrum is fitted well between $0.5-4\keV$, but the model predicts
considerably more variability at higher and lower energies than is actually
observed.

A blurred ionised reflection model was also fitted to the average spectrum in
G07.  While statistically the fit was not as good as the double power law
model, it had the advantages of describing the origin of the Fe~\ka\ emission
line and explaining the long-term (yearly) variability in a self-consistent
manner.  In adopting this spectral model for the rms simulation, the shape of 
the reflection component was kept constant throughout, but the normalisation
was allowed to vary.  Similarly the intrinsic power law was permitted to vary
in normalisation, but during the pre-dip phase its slope ($\Gamma$) was also
variable.
This model produces a very good description of the average rms spectrum
(left panel, Fig.~\ref{fig:rms}), with no significant discrepancies over
a decade of energy.
As this model provided a good description of the average rms spectrum, we 
also applied it to the rms spectra
in the four segments.  The rms spectrum in each segment was modelled
in the same way as the average rms spectrum.  The normalisations of
both continuum components were variable in all four segments, and
the photon index of the intrinsic power law was variable only in Segment A.
Interestingly, this same model
fits all four segments relatively well; in particular it describes the 
differences in Segment A compared to the other segments in a self-consistent 
manner (right panel, Fig.~\ref{fig:rms}).

Perhaps one advantage of the blurred reflection model is that its spectral
variability can be predicted to some degree.
In the light-bending scenario the reflection component and illuminating 
power law continuum vary in a particular way depending on the 
relative contribution of the two components to the total spectrum 
(Miniutti \& Fabian 2004).  As such it is of interest to examine how the
two components vary with respect to each other.

\begin{figure}
\rotatebox{270}
{\scalebox{0.32}{\includegraphics{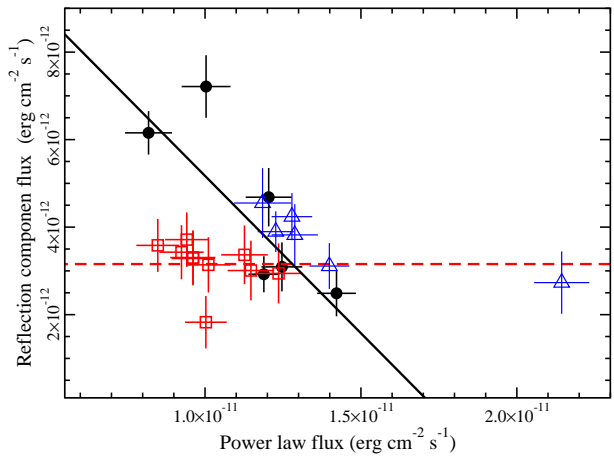}}}
\caption{The fluxes of the reflection and power law components in the $0.3-10\keV$
band are plotted against each other.  The pre-dip data (black circles) support
an anti-correlation in the relation.  Post-dip (open red squares) the reflection
component is constant despite continued fluctuations in the power law.
Block 5, which corresponds to the data at the onset of the dip (
the first block of Segment B), is shown as
a black circle in this figure ($x,y=1.0\times10^{-11},7.2\times10^{-12}$) as 
the 
behaviour is consistent with the 
Segment A data.  For comparison the 2002 data, when \iz\ was in a higher flux
state, are shown as well (blue triangles).  The relation in 2002 appear more
comparable to the pre-dip relation in 2005 than to the post-dip relation.
}
\label{fig:rpo}
\end{figure}

For each block, the $0.3-10\keV$ flux of the reflection component was plotted
against the flux of the power law component (Fig.~\ref{fig:rpo}).
As has been demonstrated regularly throughout this work the pre-dip flux-flux
relation is different from the post-dip relation.\footnote{Note that the data
from block 5 (time right at the onset of the dip) have been included in the
pre-dip relation.}  The pre-dip data show a clear anti-correlation between the
power law and reflection flux.  Whereas in the post-dip data, the reflection
component appears constant despite continued variations in the 
power law.
The different relations indicate that there is a
change in the mode of variability after the dip.

For comparison the data from fitting the 2002 observation in a similar way
have been plotted as well (open triangles in Fig.~\ref{fig:rpo}).  In
2002, when \iz\ was in a higher flux and more power law dominated state, the
flux-flux relation exhibits a negative slope and is more consistent with the 
pre-dip behaviour in 2005 (i.e. fluxes are anti-correlated).
The exception is during the 2002 flare when the single data point from that
time fell on the relation showing no correlation between the fluxes
(i.e. the post-dip relation).
This indicates that the relative amount of reflected flux increased during
that time, which is in complete agreement with the intensified
`red-wing' emission identified in figure 12 and 13 of G07.
This is consistent with the possibility that the power law emission during the
flare arises from a region physically distinct (e.g. closer to the black hole)
from the ``typical'' emitter.

\section{Discussion}
\subsection{State transitions in accreting black holes}
The 2005 observation of \iz\ reveals two distinct modes of spectral variability.
The variability in the first $\sim30\ks$ can be described with changes in 
normalisation and shape (specifically $\Gamma$) of the spectral components,
while the variability in the remaining data is attributed to fluctuations 
in normalisation alone. 

Particularly curious is that the time when the AGN makes the transition 
from one mode of spectral variability to the other is associated with
the occurrence of a flux dip in the broad-band light curve.
To the best of our knowledge, this is the first report of such ``marking''
of a mode change in an AGN.
The marking of a change in spectral behaviour with a notable event in 
the light curve may not be unique to this observation of \iz. 
Examination
of the flux-flux relation during the short 2002 \xmm\ observation 
(Fig.~\ref{fig:ff02}), along with figure 9 of G04, suggest that the
spectral change there is coincident with the modest X-ray flare.

Such state transitions, bimodal spectral variability, and flux dips and flares
are commonly seen in GBHs (e.g. Fender \& Belloni 2004;
Remillard \& McClintock 2006; Malzac \et 2006).
In addition, the behaviour shown in the right panel of Fig.~\ref{fig:hr}
is similar to that found during state changes
and oscillations in the very high state of GBHs.  
However, the rapidity of the change argues
that it is unlikely to be due to 
changes in the disc itself, but more likely related to the magnetic structure 
above the disc and could be connected with the existence of a jet.

\subsection{A jet-like nature of the corona in \iz?}
The broad iron lines observed in the 2002 and 2005 spectra of \iz\ are
very similar in appearance and were likely emitted several tens of $\rg$ from
the central black hole (G07).   Since there were no obvious time delays between
the continuum and reflection component it is reasonable to assume the
illuminating corona occupies a similar spatial region as the reflection 
component.  This suggests that at least some of the hot, optically thin plasma
blankets several tens of $\rg$ of the accretion disc and is responsible for
the mean X-ray properties exhibited by \iz.

There is also evidence indicating that the hard X-ray flare 
observed in the 2002 data originated much closer to the black hole since
the reflection spectrum during the flare appeared to be redshifted much more 
than the average broad line (G07). 
This suggests that there is a second component to the corona that is more 
compact and centrally located perhaps in a jet-like geometry.
This jet component is responsible
for the rapid events seen in \iz\ (e.g. flux flares and dips), illuminating
the inner accretion disc, and replenishing the more diffuse corona.
\iz\ is comparatively
radio quiet.  The radio source 
in \iz\ has
a steep spectrum, perhaps indicative of optically thin synchrotron emission,
and is unresolved with a physical size $\le0.38\kpc$ (Kukula
\et 1998).
The implied compactness of this proposed jet is consistent with the aborted-jet
model proposed for radio-quiet AGN (Ghisellini \et 2004).

The jet component could account for the pre-dip behaviour in \iz.
High-energy particles could be expelled via the jet, their spectrum hardening
as they propagate further from the accretion disc and source of low-energy
seed photons.  The dip itself could correspond to an interval when this ejected
component dims below the average
coronal flux (i.e. that of the extended corona).  
The scenario is not unlike that suggested to 
describe the connection between superluminal radio flares and soft
X-ray flux dips in the GBH GRS~1915+105 (Vadawale \et 2003).

The 2002 observation of \iz\ when a hard X-ray flare was detected (G04) is
plausibly consistent with this jet picture.  In that case, the 
ejected material may have collided with more material in its path resulting
in a shock and consequent flare, which then illuminated the inner accretion
disc resulting in intensified reflected emission.

While the aborted-jet model and variations along that line (e.g. Henri \& Petrucci
1997, Malzac \et 1998) can qualitatively describe our X-ray observations of \iz,
they are not unique explanations.  The compact hot plasma envisioned in \iz\ could 
be attributed to
magnetic flares atop the accretion disc (i.e. an active corona)
(e.g. Beloborodov 1999a), which could
produce similar behaviour.  There are distinctions between these two models,
some of which may be identified with future observations.

The jet model places the compact source on the spin axis of a Kerr black hole,
thus preferentially illuminating the inner accretion disc.  Such conditions
would generate extremely redshifted iron emission, similar to that seen by
Wilms \et (2001) in \mcg.  The increased photon-collection ability proposed for future 
X-ray missions can
reveal such emission lines in dimmer AGN.  In contrast,
there is no propensity for the active-corona model to produce highly 
redshifted reflection features as emission would
originate at different disc-radii.  In fact, 
Beloborodov (1999a) suggests that if the flaring plasma is composed of $e^{\pm}$ then
it should be accelerated $away$ from the disc at mildly relativistic velocities
($v/c \approx 0.1-0.7$) by radiation pressure.  This would likely
yield weaker reflection features and if the outflow is optically thick, the effect 
would be compounded as emission from the accretion disc would be obscured
(Beloborodov 1999b).  Moreover, an optically thick $e^{\pm}$ wind produces a
shifted and smeared annihilation feature at $E>511\keV$ (Beloborodov 1998), perhaps
detectable with future gamma-ray detectors.  The feature has not been detected in
the spectra of AGNs or GBHs with jets, and its presence depends on the composition of 
the jet.

Even though the escape velocity is not exceeded in the aborted-jet model, high
velocities can still be achieved.
The escape velocities from a Kerr black hole are $0.9c$ and
$0.7c$ at $2\rg$ and $4\rg$, respectively (see equation 3 of Ghisellini \et 2004).
The velocities can be much higher than what is predicted with the active-corona model.
VLBI imaging has revealed sub-pc scale jets in several radio-quiet Seyfert galaxies
(e.g. Ulvestad 2003).  It is conceivable that with improvements 
in interferometry and imaging techniques (see Ulvestad 2003 and references therein)
proper motions on even smaller scales can be measured.

The radiation scattered by the mildly relativistic wind in the active-corona model 
will have polarisation parallel to the disc normal, which may be noticeable in 
polarimetry observations.  On the other hand, beamed radiation from the jet exhibits
polarisation perpendicular to the jet.  Furthermore, the X-ray emission reflected from
an accretion disc will be polarised with angle and degree depending on conditions in 
the gravitational field, and can elucidate the geometry of the illuminating source
(e.g. Dov\v ciak, Karas \& Matt 2004).  This bolsters the demand for a 
sensitive polarimeter on future X-ray mission such as the 
X-Ray Evolving Universe Spectrometer ($XEUS$).

\subsection{On the blurred reflection interpretation}

The presence of a short jet above the black-hole and accretion-disc system
provides a possible description for the illuminating source in the 
light-bending model illustrated by Miniutti \& Fabian (2004).
In that model the accreting
black hole exists in three distinct temporal states depending on how distant 
the primary continuum source (e.g. the jet) is from the black hole.  
In the low-flux case (regime I),
the compact continuum source is within a few $\rg$ of
the black hole, and the observed spectrum is reflection dominated.  In
this case the variability of the reflection and continuum components are
correlated.  As the distance of the continuum source grows the AGN brightens and
enters regime II.  Here the variability of the reflection component reaches a 
minimum
and appears to show no correlation with the continuum changes.  The behaviour is
very similar to what we see in the post-dip state in Fig.~\ref{fig:rpo}.
As the distance of the continuum continues to increase  (e.g. $\gs20\rg$) the 
AGN enters a high-flux state and regime III.  In this state general relativistic
effects are minimal, and the reflection and continuum fluxes are anti-correlated. 
The predicted behaviour is similar to what is seen in the pre-dip state (and
possibly during the high-flux 2002 observation) (Fig.~\ref{fig:rpo}). 

Is the apparent change in behaviour marked by the flux dip in the light 
curve a transition of the black hole from regime III to II?  While the 
behaviour is consistent, the
luminosity of the continuum prior to and after the dip is not substantially
different as would be expected in the light-bending scenario.
In addition the continuum shape appears to be changing during the pre-dip 
phase.  This would indicate that if we consider the light-bending
scenario to describe the observed behaviour we must allow the compact
continuum source to undergo intrinsic changes.  This is not an unreasonable 
requirement.  If the emission process in the continuum source is Comptonisation
then one would expect that when the distance of the hot plasma (i.e. corona)
from the seed photons changes so to does the shape of the emitted spectrum.
Specifically as the distance of the corona from the seed photons increases
the emitted power law spectrum flattens.

\section{Conclusions}

The short-term spectral variability of \iz\ as observed in an $85\ks$
\xmm\ observation is discussed in detail.  The behaviour in the spectral
variability is temporally distinct and marked by a dip in the broad-band
count rate at $\sim30\ks$ into the observation.  Prior to the dip 
the variability requires changes in shape and normalisation of the 
spectral components.  Only changes in normalisation are required after the dip.
The changes occur on dynamically short timescales and likely would not be
observable in GBHs.
 
The ionised reflection model is applied to the temporal data with good
agreement.  The interpretation can follow the light-bending scenario, but
requires intrinsic changes in the shape of the primary continuum source prior
to the dip.  If we consider the corona to be the base of a possible
jet then the spectral changes could arise from material that is
expelled from it.  The aborted jet scenario seems consistent with the
2002 and 2005 observations of \iz, which possesses parallels with some GBH
observations.


\section*{Acknowledgements}

The \xmm\ project is an ESA Science Mission with instruments
and contributions directly funded by ESA Member States and the
USA (NASA). The \xmm\ project is supported by the
Bundesministerium f\"ur Wirtschaft und Technologie/Deutsches Zentrum
f\"ur Luft- und Raumfahrt (BMWI/DLR, FKZ 50 OX 0001), the Max-Planck
Society and the Heidenhain-Stiftung.
For helpful suggestions and insightful conversations LCG thanks
Andrea Merloni, Jon Miller and Iossif Papadakis.
WNB acknowledges support from NASA LTSA grant NAG5-13035 and NASA grant NNG05GR05G.



\bsp
\label{lastpage}

\begin{thebibliography}{}

 

\bibitem{}
Belloni T., Mendez M., King A. R., van der Klis M., van Paradijs J., 1997,
ApJ, 479, L145

\bibitem{}
Beloborodov A. M., 1998, ApJ, 496, L105

\bibitem{}
Beloborodov A. M., 1999a, ApJ, 510, L123

\bibitem{}
Beloborodov A. M., 1999b, MNRAS, 305, 181 

\bibitem{}
Boller Th., Brandt W. N., Fabian A. C., Fink H. H., 1997, MNRAS, 289, 393

\bibitem{}
Brandt W. N., Boller Th., Fabian A. C., Ruszkowski M. 1999, MNRAS, 303, L58


\bibitem{}
Costantini E., Gallo L. C., Brandt W. N., Fabian A. C., Boller Th.,
2007, accepted by MNRAS (astro-ph/0702553) 

\bibitem{}
Dov\v ciak M., Karas V., Matt G., 2004, MNRAS, 355, 1005

\bibitem{}
Edelson R., Turner T. J., Pounds K., Vaughan S. Markowitz A.,
Marshall H., Dobbie P., Warwick R., 2002, ApJ, 568, 610

\bibitem{}
Esin A. A., Narayan R., Cui W., Grove J. E., Zhang S., 1998,
ApJ, 505, 854

\bibitem{}
Fabian A. C. \et, 2002, MNRAS, 335, L1

\bibitem{}
Fabian A. C., Miniutti G., Gallo L., Boller Th., Tanaka Y., Vaughan S., 
Ross R. R., 2004, MNRAS, 353, 1071

\bibitem{}
Fender R., Belloni T., 2004, ARA\&A, 42, 317


\bibitem{}
Gallo E., Fender R. P., Pooley G. G., 2003, MNRAS, 344, 60

\bibitem{}
Gallo L. C., Boller Th., Brandt W. N., Fabian A. C., Vaughan S., 2004a, A\&A,
417, 29 (G04)

\bibitem{}
Gallo L. C., Boller Th., Tanaka Y., Fabian A. C., Brandt W. N., Welsh W. F.,
Anabuki N., Haba Y., 2004b, MNRAS, 347, 269

\bibitem{}
Gallo L. C., Tanaka Y., Boller Th., Fabian A. C., Vaughan S., Brandt W. N., 
2004c, MNRAS, 353, 1064 

\bibitem{}
Gallo L. C., Brandt W. N., Costantini E. Fabian A. C., Iwasawa K., 
Papadakis I. E., 2007, accepted by MNRAS (astro-ph/0610283) (G07)

\bibitem{}
Gierlinski M., Done C., 2006, MNRAS, 371, 16 

\bibitem{}
Gilfanov M., Revnivtsev M., Molkov S., 2003, A\&A, 410, 217

\bibitem{}
Greiner J., Morgan E. H., Remillard R. A., 1996, ApJ, 473, L107

\bibitem{}
Ghisellini G., Haardt F., Matt G., 2004, A\&A, 413, 535

\bibitem{}
Henri G., Petrucci P. O., 1997, A\&A, 326, 87

\bibitem{} Jansen F. \et 2001, A\&A, 365, L1 

\bibitem{}
Malzac J. G. \et, 2006, A\&A, 448, 1125

\bibitem{}
Malzac J., Jourdain E., Petrucci P. O., Henri G., 1998, A\&A, 336, 807

\bibitem{}
Markoff S., Falcke H., Fender R., 2001, A\&A, 372, L25

\bibitem{}
Markowitz A., Papadakis I. E., Arevalo P., Turner T. J., Miller L., 
Reeves J. N., 2006, ApJ accepted (astro-ph/0611072)

\bibitem{}
McClintock J. E., Remillard R. A., 2006, in Lewin W. H. G., van der Klis M., 
eds, {\it Compact Stellar X-Ray Sources}. Cambridge Univ. Press, 
(astro-ph/0306213)

\bibitem{}
McHardy I. M., Koerding E., Knigge C., Uttley P., Fender R. P., 2006, Nature,
444, 730

\bibitem{}
Merloni A., Heinz S., di Matteo T., 2003, MNRAS, 345, 1057

\bibitem{}
Miller J. M., Homan J., Steeghs D., Rupen M., Hunstead R. W., Wijnands R., 
Charles P. A., Fabian A. C., 2006, ApJ, 653, 525

\bibitem{}
Miniutti G., Fabian A. C., 2004, MNRAS, 349, 1435

\bibitem{}
Ponti G., Cappi M., Dadina M., Malaguti G., 2004, A\&A, 417, 451

\bibitem{}
Kukula M. J., Dunlop J. S., Hughes D. H., Rawlings S., 1998, MNRAS, 297, 366

\bibitem{}
Remillard R. A., McClintock J. E., 2006, ARA\&A, 44, 49

\bibitem{}
Shakura N. I., Sunyaev R. A., 1973, A\&A, 24, 337

\bibitem{}
Taylor R., Uttley P., McHardy I., 2003, MNRAS, 342, 31

\bibitem{}
Ulvestad J. S., 2003, in Zensus J. A., Cohen M. H., Ros E., 
eds, {\it Radio Astronomy at the Fringe}. ASP Conf. Series Vol. 300, p97

\bibitem{}
Vadawale S. V., Rao A. R., Naik S., Yadav J. S., Ishwara-Chandra C. H., 
Pramesh Rao A., Pooley G. G., 2003, ApJ, 597, 1023

\bibitem{}
Wilms, J., Reynolds C. S., Begelman M. C., Reeves J., Molendi S., Staubert R., 
Kendziorra E., 2001, MNRAS, 328, L27


\bibitem{}
Vestergaard M., Peterson B., 2006, ApJ, 641, 689

\bibitem{}
Zdziarski A. A., 2005, MNRAS, 360, 816

\end{thebibliography}
\end{document}